\documentclass[10pt, conference]{IEEEtran}
\usepackage{cite}
\usepackage{amsmath,amssymb,amsfonts}
\usepackage{algorithmic}
\usepackage{graphicx}
\usepackage{textcomp}
\usepackage{xcolor}

\usepackage[colorlinks,linkcolor=black,anchorcolor=black,citecolor=purple]{hyperref}
\usepackage{listings}
\usepackage{multirow}
\usepackage{subfigure}
\usepackage{wrapfig}
\usepackage{booktabs}
\usepackage[T1]{fontenc}
\usepackage{algorithm}
\usepackage{threeparttable}
\usepackage{tikz}
\usetikzlibrary{calc}
\usepackage{pgfplots}
\usepackage{mdframed}
\usepackage{xspace}

\def\BibTeX{{\rm B\kern-.05em{\sc i\kern-.025em b}\kern-.08em
    T\kern-.1667em\lower.7ex\hbox{E}\kern-.125emX}}

\newcommand{\CodeIn}[1]{{\small \texttt{#1}}}

\newcommand{\MYCOMMENT}[1]{}
\newcommand{\Space}[1]{}

\newcommand{\tool}{\textit{ConfInLog}\xspace}

\definecolor{lightgreen}{rgb}{0.90, 1, 0.90}
\definecolor{lightred}{rgb}{1, 0.90, 0.90}
\definecolor{grey}{RGB}{88, 87, 86}

\lstdefinelanguage{myC}[]{C}{
  morekeywords={uint},
  morecomment=[f][\color{blue}]{@@},     
  morecomment=[f][\color{red}]-,         
  morecomment=[f][\color{ForestGreen}]+, 
  morecomment=[f][\color{magenta}]{---}, 
  morecomment=[f][\color{magenta}]{+++},
  morecomment=[f][\color{magenta}]{//},
  morecomment=[f][\color{magenta}]{/*},
  morecomment=[f][\color{magenta}]{\ \ \ \ //},
  morecomment=[f][\color{magenta}]{\ \ //},
  keywordstyle=\color{blue},
  commentstyle=\color{gray},
  stringstyle=\color{purple},
  numbers=left,
  basicstyle={\scriptsize\ttfamily},
  numbersep=6pt,
  numberstyle=\tiny\ttfamily,
  breaklines=true,
  escapeinside={(*@}{@*)},
  showstringspaces=false,
  tabsize=2,
  frame=tblr,
  xleftmargin=8pt,
}

\lstdefinelanguage{myCinline}[]{C}{
  morekeywords={uint},
  morecomment=[f][\color{blue}]{@@},     
  morecomment=[f][\color{red}]-,         
  morecomment=[f][\color{ForestGreen}]+, 
  morecomment=[f][\color{magenta}]{---}, 
  morecomment=[f][\color{magenta}]{+++},
  morecomment=[f][\color{magenta}]{//},
  morecomment=[f][\color{magenta}]{/*},
  morecomment=[f][\color{magenta}]{\ \ \ \ //},
  morecomment=[f][\color{magenta}]{\ \ //},
  keywordstyle=\color{blue},
  commentstyle=\color{gray},
  stringstyle=\color{purple},
  numbers=left,
  basicstyle={\scriptsize\ttfamily},
  numbersep=6pt,
  numberstyle=\tiny\ttfamily,
  breaklines=true,
  escapeinside={(*@}{@*)},
  showstringspaces=false,
  tabsize=2,
  xleftmargin=8pt,
}

\begin{document}

\title{ConfInLog: Leveraging Software Logs to Infer Configuration Constraints\\
}

\author{
\IEEEauthorblockN{Shulin Zhou, Xiaodong Liu, Shanshan Li, Zhouyang Jia, \\Yuanliang Zhang, Teng Wang, Wang Li, Xiangke Liao}
\IEEEauthorblockA{National University of Defense Technology, Changsha, China\\
\{zhoushulin, liuxiaodong, shanshanli, jiazhouyang\}@nudt.edu.cn\\
\{zhangyuanliang13, wangteng13, liwang2015, xkliao\}@nudt.edu.cn}
}

\maketitle

\begin{abstract}
Misconfigurations have become the dominant causes of software failures in recent years, drawing tremendous attention for their increasing prevalence and severity. 
Configuration constraints can preemptively avoid misconfiguration by defining the conditions that configuration options should satisfy.
Documentation is the main source of configuration constraints, but it might be incomplete or inconsistent with the source code.
In this regard, prior researches have focused on obtaining configuration constraints from software source code through static analysis. 
However, the difficulty in pointer analysis and context comprehension prevents them from collecting accurate and comprehensive constraints. 
In this paper, we observed that software logs often contain configuration constraints.
We conducted an empirical study and summarized patterns of configuration-related log messages.
Guided by the study, we designed and implemented \tool, a static tool to infer configuration constraints from log messages. 
\tool first selects configuration-related log messages from source code by using the summarized patterns, 
then infers constraints from log messages based on the summarized natural language patterns.
To evaluate the effectiveness of \tool, we applied our tool on seven popular open-source software systems.
\tool successfully inferred 22$\sim$163 constraints, in which 59.5\%$\sim$61.6\% could not be inferred by the state-of-the-art work. 
Finally, we submitted 67 documentation patches regarding the constraints inferred by \tool. The constraints in 29 patches have been confirmed by the developers, among which 10 patches have been accepted.
\end{abstract}


\section{Introduction}
\label{sec:intro}

In recent years, misconfigurations have become one of the dominant causes of software failures, 
drawing tremendous attention for their increasing prevalence and severity\cite{barroso2018datacenter,tang2015holistic,maurer2015fail,facebook2019outagenews,google2019networkissue,spacex2020destory,amvrosiadis2016getting,gunawi2016does,jiang2008disks,kendrick2012takes,rabkin2013how}. 
One major reason for the prevalent misconfiguration issues is the ever-increasing complexity of configuration\cite{xu2015hey,liao2018you,nadi2014mining,rabkin2011static,santolucito2017synthesizing}. 
It is challenging for novices or even proficients to correctly understand the constraints of configuration options without expert domain knowledge\cite{yin2011empirical,xu2015hey}. 
To preemptively avoid misconfiguration, 
documentation is the main source for users to comprehend configuration constraints.
However, as mentioned by \cite{rabkin2011static,xu2013not,liao2018you}, 
documentation is often incomplete or outdated, making it difficult for users to learn configuration constraints. 
For instance, the value range (\textit{0$\sim$32767}) of configuration option "\textit{LimitRequestFields}" in Apache Httpd\cite{httpdhomepage} is a legacy mistake in documentation\cite{httpdbug64893} . 

\begin{figure}[tb]
	\centering
	\includegraphics[width=.5\textwidth]{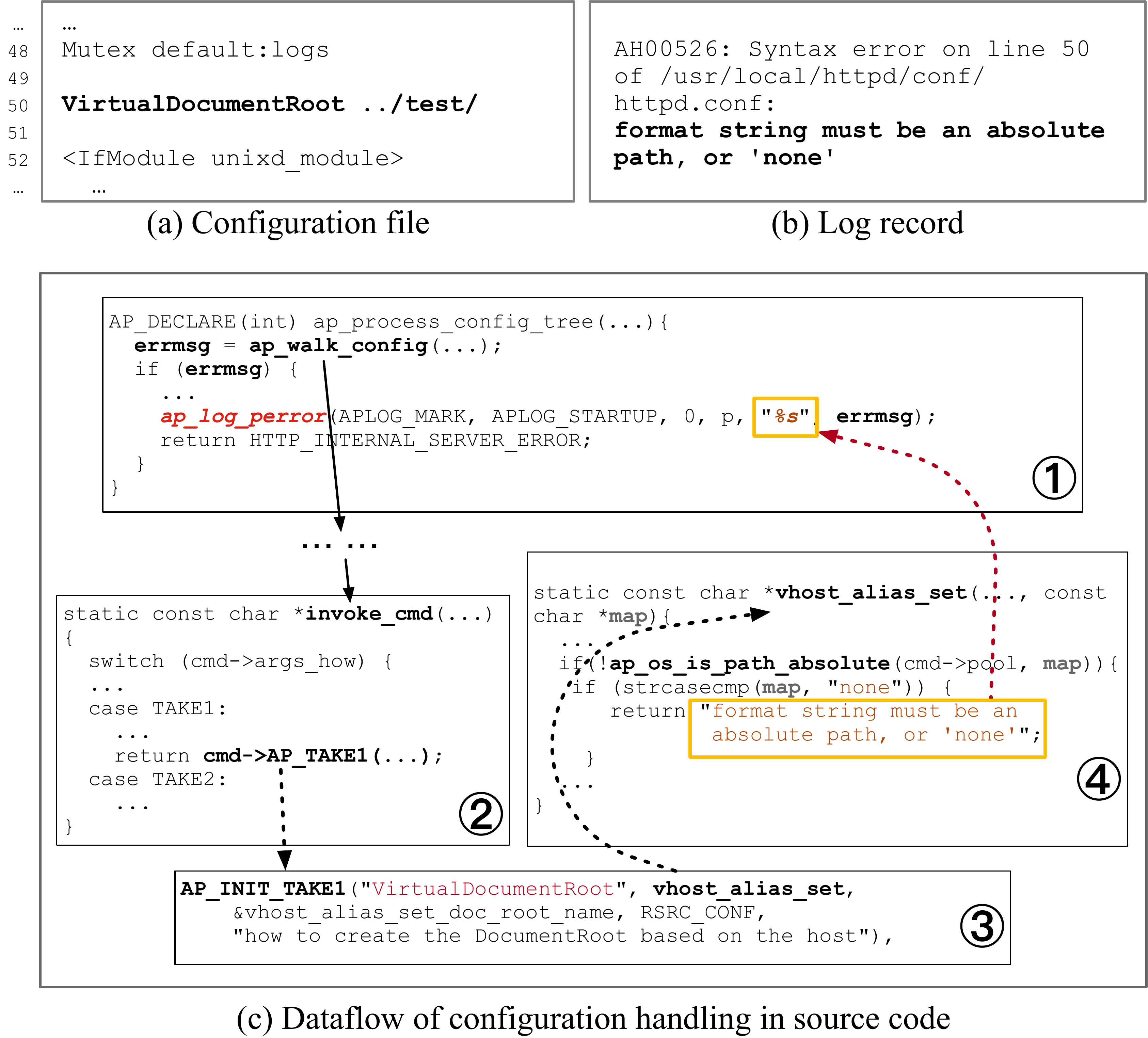}
	\vspace{-20pt}
	\caption{A motivated example of configuration constraint in log.}
	\label{fig:motivating-example}
	\vspace{-10pt}
\end{figure}

In order to help users understand the configuration constraints, 
there are prior methods\cite{xu2013not,liao2018you,chen2020understanding} focusing on inferring configuration constraints from software source code. 
The existing methods typically have two main steps.
First, they map configuration options to program variables, 
and then track usages of the variables by using static program analysis.
Second, they infer constraints when the variables occur
in some predefined code patterns.
For example, a typical pattern is to extract constraints from conditional expressions of branch statements, which contain the usages of the variables. 

This process, however, may be limited in practice.
On one hand, the pointer mechanism in programming languages like C/C++ may bring great barriers in static program analysis,
especially for large-scale software systems,
limiting the analysis scope and accuracy\cite{Landi1991Pointer}. 
On the other hand, the quality of these predefined patterns is highly depended on personal experience.
As a result, they may miss the constraints that are hard to be tracked by static analysis, or miss the constraints that do not belong to any pattern.

In this paper, we observe that the \textit{log records} (referring to the text outputs at runtime) of software systems usually contain constraints of configuration options, as exemplified in Fig.\ref{fig:motivating-example}. 
In Fig.\ref{fig:motivating-example}(a), 
the configuration option "VirtualDocumentRoot" is set to "\textit{../test/}" in the configuration file of Apache Httpd; 
Fig.1(b) presents the log record after running the official test suite, 
indicating that "\textit{VirtualDocumentRoot}" should be an absolute path or "none". 
This constraint is clearly defined in the \CodeIn{Return} statement in code snippet \textcircled{\scriptsize 4} of Fig.\ref{fig:motivating-example}(c).
This \CodeIn{Return} statement is controlled by two branch statements.
However, it is difficult to infer constraint from the conditions of the branch statements by using the traditional static analysis method, since it is hard to understand the semantics of function \CodeIn{ap\_os\_is\_path\_absolute()}, and the comparison between \CodeIn{map} and "none" does not match any predefined code patterns.

To figure out the prevalence of configuration constraints in logs, 
we carried out an empirical study on four popular open-source software systems.
The result shows that 57.0\% (85/149) of configuration options have constraint descriptions in log records.
By tracing log records back to source code, 
we found that \textit{log messages} (referring to the string constants in source code that will be printed by log statements) always indicate the key information of constraints in log records. 
Yet it is non-trivial to identify whether a log message is related to a configuration option, 
since log messages of 77.6\% (66/85) configuration options do not contain the option names.
With a further analysis, we summarized patterns which could be used to effectively identify configuration-related log messages. 

Guided by our findings, we proposed \tool, 
a practicable tool to statically infer configuration constraints from log messages in source code. 
\tool first selects configuration-related log messages from source code
through matching the patterns of how log messages relate to configuration options.
Then, \tool infers the constraints from configuration-related log messages based on natural language patterns summarized from existing constraints descriptions.
An experiment on seven popular open-source software systems proved the effectiveness of \tool. 
\tool could infer 22$\sim$163 constraints from the software systems, 
and 59.5\%$\sim$61.6\% of them could not be inferred by the state-of-the-art work.
Also, after manually checking the documentation of those software systems, 
we submitted 67 documentation patches regarding the constraints inferred by \tool.
The constraints in 29 patches have been confirmed by the developers, among which 10 patches have been accepted.
The others are still pending for responses at the time of writing.

In summary, this paper makes the following contributions:

\begin{itemize}
\item We observed that logs contain configuration constraints based on an empirical study on four popular open-source software systems, and summarized patterns to select the configuration-related log messages.
\item We designed and implemented \tool, an automated tool to infer configuration constraints from log messages in source code, which could be an effective complement to previous work with static program analysis.
\item We evaluated \tool on seven open-source software systems. \tool could infer 22$\sim$163 configuration constraints, in which 59.5\%$\sim$61.6\% could not be inferred by the state-of-the-art work. We submitted 67 patches based on our inferred constraints. The constraints in 29 patches have been confirmed by the developers, among which 10 patches have been accepted.
\end{itemize}

The remainder of this paper is organized as follows. In Section \ref{sec:study}, we present the empirical study. In Section \ref{sec:design}, we first provide an overview of \tool, and then describe the details of each step. In Section \ref{sec:evaluation}, we elaborate the experimental setup and report the experimental results. In Section \ref{sec:threats}, we discuss the threats to validity. In Section \ref{sec:rw}, we briefly review the related work. Finally, we conclude our study and future work in \ref{sec:conclusion}.

\section{Empirical Study}
\label{sec:study}

\newcounter{findingcounter}

In this section, we take a in-depth look into the configuration constraints in logs through an empirical study. We manually analyzed the prevalence and characteristics of constraint-related logs. In our observation, we found that configuration constraints are widely existed in the log records of software systems, and there are characteristics to select log messages related to configuration options.

\subsection{Prevalence of Configuration Constraints in Logs}
\label{sec:study:result}

\begin{figure}[tb]
	\centering
	\includegraphics[width=.5\textwidth]{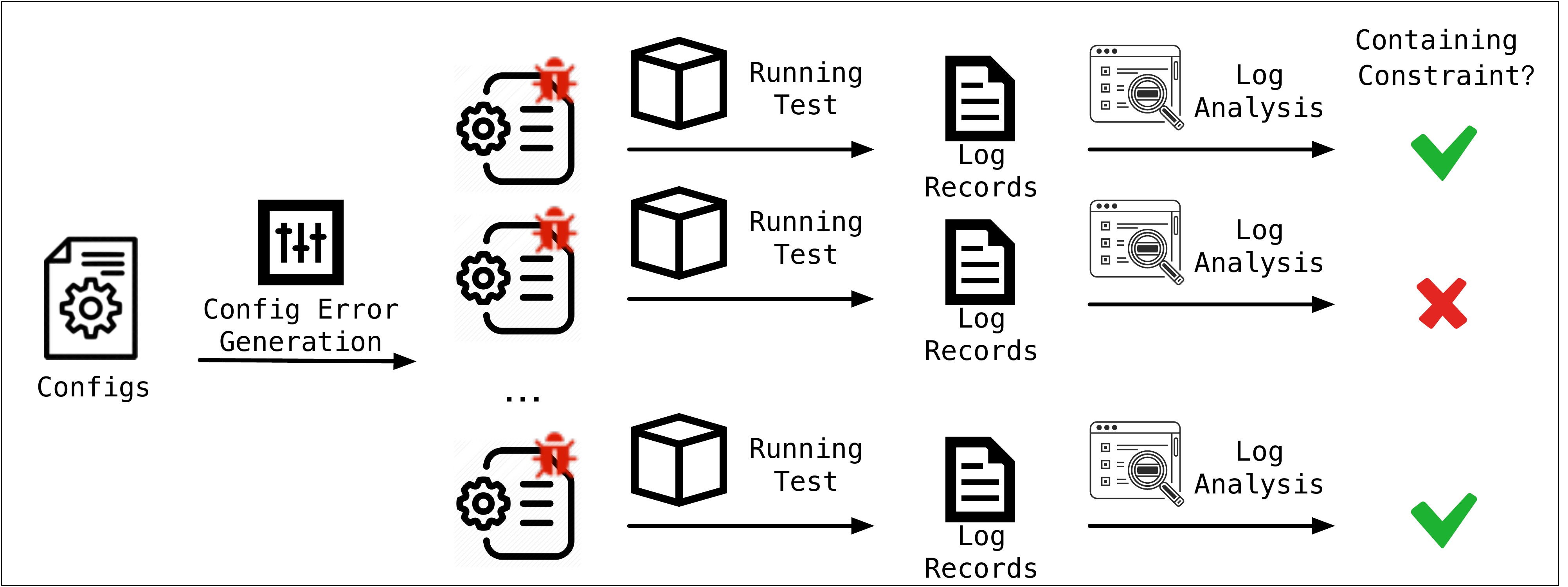}
	\vspace{-20pt}
	\caption{Workflow of Configuration Error Injection.}
	\label{fig:misconfig-inject}
	\vspace{-10pt}
\end{figure}

To acquire log records from software systems, 
we chose to inject configuration errors, 
as the workflow shown in Fig.\ref{fig:misconfig-inject}.
There are mainly two benefits. 
It is not necessary to comprehend the details of code logics, 
and it is more efficient to select the log records related to configuration options.

\begin{table*}[tb]
	\caption{Proportion of Configuration Options with Different Reaction in Log Records.}
	\label{table:log-proportion-in-misconfig-injection}
	\centering
	\setlength{\tabcolsep}{6mm}{
		\renewcommand{\arraystretch}{1.2}
		\begin{tabular}{ccccc}
			\toprule
			\multirow{2}{*}{\textbf{Software}} &
			\multirow{2}{*}{\textbf{\# of Config Options Under Test}} &
			\multicolumn{3}{c}{\textbf{\# of Config Options with Related Logs}} \\ \cline{3-5} 
				&	& \textbf{Total} & \textbf{with Constraints} & \textbf{without Constraints} \\
			\midrule
				Httpd		& 	46	&	43	&	34		& 9		\\
				Nginx		& 	28	& 	25	& 	9		& 16	\\
				MySQL		&   23	&	23	&	11		& 12	\\
				PostgreSQL	& 	58	&	58	&	31		& 27	\\ 
			\midrule
				\textbf{Total} & \textbf{155} &	\textbf{149}	& \textbf{85}	& \textbf{64}	\\ 
			\bottomrule
		\end{tabular}
	}
\end{table*}

\textbf{\romannumeral1) Data Set:} As shown in Table \ref{table:log-proportion-in-misconfig-injection}, we choose four well-known software systems for our study. All of these software systems are open-source server software and widely deployed in the field, and provide official test suite to run different workloads.
Considering the large number of configuration options provided by these software systems, we choose the core modules or functions of each system to conduct configuration error injection. For Httpd and Nginx, we focus on the configuration options in Core modules; for MySQL, we focus on the configuration options related to MyISAM storage module; for PostgreSQL, we test the configuration options related to file locations, connections, authentication, and resource consumption.

\textbf{\romannumeral2) Configuration Error Injection:} Prior studies\cite{li2018confvd,zhang2015proactive,xu2013not,arshad2013characterizing,keller2008conferr} on configuration error generation, mainly fall into two categories, namely mutation based\cite{keller2008conferr,arshad2013characterizing,zhang2015proactive} and constraint (specification) violation based \cite{xu2013not,li2018confvd}.
Considering that the constraint violation based methods require the configuration constraints beforehand, 
we choose to implement the mutation rules in the state-of-the-art mutation-based method \cite{zhang2015proactive} to generate configuration errors in our study. Then we inject one configuration error each time, and run the official test suite to collect relative log records.

\textbf{\romannumeral3) Log Analysis:} After running the test suites, there are plenty of log records generated by the software system under test. To figure out whether there is configuration constraint in log record, we apply a semi-automatically approach to analyze the log records. 
First, for a configuration option under test, we preprocess the configuration option for stemming and tokenization to a word set $\psi$. 
If there is any match of a) word in $\psi$, 
b) the configuration file name, 
or c) keywords "config", "parameter", "directive",  with the log records, 
we manually inspect the log records to check whether there is any constraint description of the configuration option under test. 
We regard a log record as related to the option under test if it satisfies one of the following conditions:
a) the log record contains the name of the option under test; 
b) it contains all the words in the option connected with common delimiters; 
c) it contains the location information (line number) to find the option in configuration file.
Moreover, if the log record is related to the option under test, and provides guidance for fixing the configuration error, we regard it as containing configuration constraints.

\addtocounter{findingcounter}{1}
\begin{mdframed}[everyline=true]
	\textbf{Finding \arabic{findingcounter}}: \textit{Up to 57.0\% of the configuration options have log records containing constraints description, which provides opportunity to infer configuration constraints.}
\end{mdframed}

As shown in Table \ref{table:log-proportion-in-misconfig-injection}, 
we tested a total of 155 configuration options, 
among which 96.1\% (149/155) have related log records. 
More importantly, 57.0\% (85/149) have explicit constraints description in the log records. 
For those options without constraints in the log records (64/149), 
7.8\% (5/64) have descriptions about the purpose of configuration options which, 
however, are not enough to guide the fixing. 
For example, the log records of configuration option "\textit{UseCanonicalName}" indicate that it defines "\textit{how to work out the ServerName : Port when constructing URLs}", 
but there is no description about what value should be set.
92.2\% (59/64) have log records describing the error status without further guidance of error fixing. 
For example, configuration option "\textit{worker\_aio\_requests}" in Nginx only outputted "invalid value" in the related log records.

\subsection{Analysis of Constraints-Related Logs}
\label{sec:study:analysis}

In light of the wide existence of configuration constraints in log records, we hold the hypothesis that we could infer configuration constraints from logs. Nevertheless, it is challenging to trigger all the log records, considering the various workloads and the limited capability of current methods for generating configuration errors (we will discuss the circumstances in Section \ref{sec:evaluate:RQ3:dynamic}).
Consequently, alternative static analysis strategies are taken into consideration. We took a further look at the logs in source code.

\subsubsection{\textbf{The Existing Styles of Log Messages}}
\label{sec:study:analysis:form}

To infer configuration constraints from logs in the source code, 
a straightforward method is to locate the \textit{log statements} (referring to the program statements in the source code that perform the log behaviors), 
and check whether there are possible configuration constraints.
However, this approach is not feasible in practice. 
First, as mentioned by \cite{jia2018smartlog}, the huge number of log functions in large-scale software and the collaborative development make it a non-trivial task to recognize all the log statements in source code. 
Second, even if we could locate all the log statements, 
it is still difficult to obtain the corresponding log message. 
Take Fig.\ref{fig:motivating-example}(c) as example, 
the log message is not explicitly presented in the log statement \CodeIn{ap\_log\_perror()} in code snippet \textcircled{\scriptsize 1}, 
but determined by the return value \CodeIn{errmsg} of function \CodeIn{ap\_walk\_config()}, 
which is indeterministic due to the invocations of different function pointers in function \CodeIn{invoke\_cmd()} in code snippet \textcircled{\scriptsize 3}, 
and the outputted log message is existed in the \CodeIn{Return} Statement in code snippet \textcircled{\scriptsize 4}.

\begin{table*}[tb]
	\caption{Proportion of Different Existing Styles of Log Messages in Source Code.}
	\label{table:styles-of-log-message}
	\centering
	\setlength{\tabcolsep}{6mm}{
		\begin{tabular}{cccccc}
		\toprule
		\textbf{Software} &
		\textbf{\begin{tabular}[c]{@{}c@{}}\# of Sampled \\ Log Stmt\end{tabular}} &
		\textbf{\begin{tabular}[c]{@{}c@{}}\# of Log Msg\\ in Log Stmt\end{tabular}} &
		\textbf{\begin{tabular}[c]{@{}c@{}}\# of Log Msg\\ in Return Stmt\end{tabular}} &
		\textbf{\begin{tabular}[c]{@{}c@{}}\# of Log Msg\\ in Assign Stmt\end{tabular}} &
		\textbf{\begin{tabular}[c]{@{}c@{}}\# of Log Msg\\ in Structure\end{tabular}} \\ 
		\midrule
			Httpd		& 100	& 78	& 16	& 3		& 3		\\
			Nginx		& 100	& 82	& 8		& 10	& 0		\\
			MySQL		& 100	& 93	& 3		& 4		& 0		\\
			PostgreSQL	& 100	& 97	& 0		& 3		& 0		\\ 
		\midrule
			\textbf{Total} & \textbf{400} & \textbf{350} & \textbf{27} & \textbf{20} & \textbf{3} \\ 
		\bottomrule
		\end{tabular}
	}
\end{table*}

\addtocounter{findingcounter}{1}
\begin{mdframed}[everyline=true]
	\textbf{Finding \arabic{findingcounter}}:  \textit{
	12.5\% of the log messages are not explicitly presented in the log statements, but exist in various styles with implicit dataflow to log statements.}.
\end{mdframed}

\begin{figure}[tb]
	\centering
	\includegraphics[width=.5\textwidth]{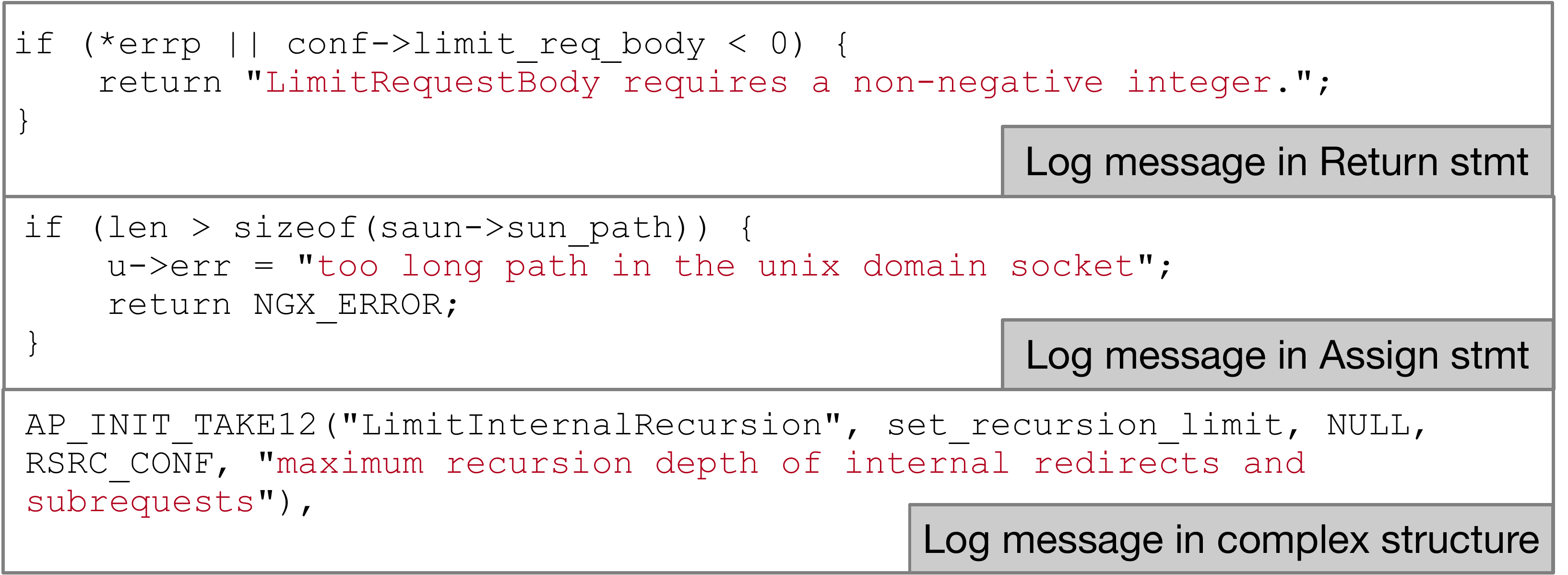}
	\vspace{-20pt}
	\caption{Different existing styles of log messages in source code.}
	\label{fig:log-message-styles}
	\vspace{-10pt}
\end{figure}

In order to explore the characteristics of log messages in the source code,
we randomly sampled 100 log statements in source code for each studied software system, 
and manually inspected the context and dataflow to find the corresponding log messages. 
As shown in Table \ref{table:styles-of-log-message}, 
except for the 87.5\% (350/400) of log statements that contain log messages directly inside the statement, 
the remaining 12.5\% (50/400) are implicitly existed in three different styles, 
as the code examples shown in Fig.\ref{fig:log-message-styles}.
Besides, due to the usage of function pointers, 
a log statement may be reused to output different log messages by calling different function pointers, 
as the motivating example in Fig.\ref{fig:motivating-example} shown. 
These circumstances all prevent us from collecting all the log messages directly from log statements.

\vspace{3pt}
\subsubsection{\textbf{Identification of Configuration-related Log Messages}}
\label{sec:study:analysis:relationship}

\addtocounter{findingcounter}{1}
\begin{mdframed}[everyline=true]
	\textbf{Finding \arabic{findingcounter}}:  \textit{Up to 77.6\% of log messages do not directly specify the the configuration option involved, which limits the identification of configuration-related log messages.}
\end{mdframed}

To infer configuration constraints from log messages, 
we need to identify whether the log message is related to specific configuration option at first.
Nevertheless, a simple keyword matching for options' name is far from enough.
Among all the 85 configuration options with constraints in log records found in Section \ref{sec:study:result}, 
up to 77.6\% of the configuration-related log messages cannot be identified by directly matching the options' name.

\addtocounter{findingcounter}{1}
\begin{mdframed}[everyline=true]
	\textbf{Finding \arabic{findingcounter}}:  \textit{51.8\% of configuration-related log messages could be identified by directly matching option names in log messages, or finding relevant functions or variables, which implies the feasibility of identifying configuration-related log messages}.
\end{mdframed}

\begin{figure}
	\begin{minipage}{0.99\linewidth}
\begin{lstlisting}[language=myC]
if (clcf->(*@\colorbox{lightred}{disable\_symlinks}@*) == NGX_CONF_UNSET_UINT) {
    ngx_conf_log_error(NGX_LOG_EMERG, cf, 0,
        "\"%V\" must have \"off\", \"on\" "
        "or \"if_not_owner\" parameter",
        &cmd->name);
    return NGX_CONF_ERROR;
}
\end{lstlisting}
		\vspace{-3.5pt}
		\centerline{(a) The log message is influenced by variable}
		\centerline{related to configuration option.}
		\vspace{3.5pt}
	\end{minipage}
	\begin{minipage}{0.99\linewidth}
\begin{lstlisting}[language=myC]
AP_INIT_TAKE1("(*@\colorbox{lightgreen}{AllowEncodedSlashes}@*)", (*@\colorbox{lightred}{set\_allow2f}@*),NULL,RSRC_CONF,
   "Allow URLs containing '/' encoded as '%2F'"),

static const char *(*@\colorbox{lightred}{set\_allow2f}@*)(cmd_parms *cmd, void *d_,
                               const char *arg)
{
   if (0 == strcasecmp(arg, "on")) {
      d->allow_encoded_slashes = 1;
      d->decode_encoded_slashes = 1;
   }
   ...
   else {
      return apr_pstrcat(cmd->pool,
               cmd->cmd->name, " must be On, Off, or NoDecode",
               NULL);
   }
}
\end{lstlisting}
		\vspace{-3.5pt}
		\centerline{(b) The log message is influenced by function}
		\centerline{related to configuration option.}
		\vspace{3.5pt}
	\end{minipage}
    \vspace{-10pt}
    \caption{Different patterns of log messages related to configuration options.}
	\label{fig:different-kinds-of-match-types}
	\vspace{-10pt}
\end{figure}

With a further manual analysis in source code, we found that, 
except for the 22.4\%(19/85) of log messages that directly contain name information inside, 
29.4\%(25/85) of log messages fall into two code patterns in Fig.\ref{fig:different-kinds-of-match-types}, which are helpful in identifying the configuration-related log messages without name information. 

In the first pattern, the log message is influenced by the program variable related to specific configuration option. 
Take Fig.\ref{fig:different-kinds-of-match-types}(a) as example, 
the log message is under control of program variable \CodeIn{clcf->disable\_symlinks} in the branch condition, 
which stores the value of the configuration option "\textit{disable\_symlinks}".
In the second pattern, the log message is influenced by the function related to specific configuration option. 
Fig\ref{fig:different-kinds-of-match-types}(b) shows an example, 
the log message is controlled by the argument \CodeIn{arg} of function \CodeIn{set\_allow2f()} in \CodeIn{If/Else} branch statement, 
while this function is related to configuration option "\textit{AllowEncodedSlashes}" with definition in the structure of \CodeIn{AP\_INIT\_TAKE1} macro.

\begin{figure*}[htb]
	\centering
	\includegraphics[width=.9\textwidth]{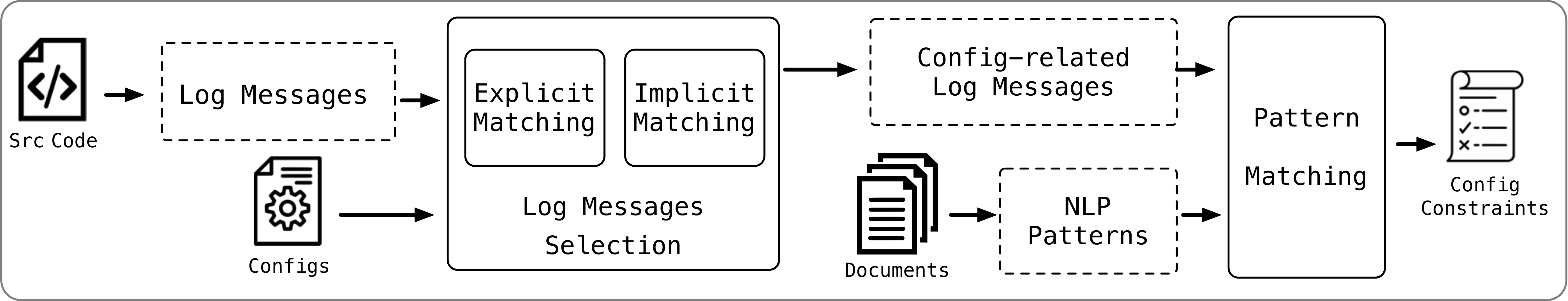}
	\vspace{-10pt}
	\caption{Workflow of \tool.}
	\label{fig:logcons-overview}
	\vspace{-10pt}
\end{figure*}

For the remaining 48.2\% (41/85) of configuration options that could not identify the related log messages based on finding \arabic{findingcounter}, all of them are numeric or boolean types, and handled by general logic with predefined information in specific structures. For instance, the configuration options with integer value in PostgreSQL are checked by the \CodeIn{min/max} bound defined in struct \CodeIn{config\_int}. These constraints could be easily learned in prior work\cite{liao2018you}, without the need to infer from log messages.

\section{Design and Implementation}
\label{sec:design}

In this section, we present the design and implementation of \tool. 
Fig.\ref{fig:logcons-overview} illustrates the workflow.
\tool takes source code and configuration options as input, 
and outputs a list of configuration constraints corresponding to different options. 
\tool consists of two major steps: Log messages selection and constraints inferring.
First, \tool collects the log messages in target software and selects the configuration-related ones.
Then, based on summarized NLP patterns, \tool infers the configuration constraints from selected log messages.

\subsection{Log Messages Selection}
\label{sec:design:select}

As mentioned in Section \ref{sec:study:analysis:form}, 
not all the log messages are explicitly present in the log statements. 
In order to get as much log messages as possible, 
\tool considers all the string constants in the code as candidate log messages.
The false positive messages will be filtered out in later steps. 
Since a log message may be spliced by several string constants, variables and macros,
\tool first converts the variable to a special label "\textit{\_VARIABLE\_}" and replaces the constant and macros, then joins them together. 
As shown in the following code snippet, 
the log message that \tool collects is "\textit{LimitRequestFields \_VARIABLE\_ must be a non-negative integer (0 = no limit)}".

\begin{lstlisting}[language=myCinline]
apr_pstrcat(cmd->temp_pool, "LimitRequestFields \"", arg,
	"\" must be a non-negative integer (0 = no limit)", NULL);
\end{lstlisting}

After collecting candidate log messages, \tool takes a two-step approach to select configuration-related log messages.

\subsubsection{\textbf{Finding Configuration Options in Log Messages}}
\label{sec:design:select:direct}

The direct method to identify configuration-related log messages is to check whether the log messages contain option name information inside.
First, \tool splits each configuration option into words, and generates a regular expression with possible delimiters between the words. 
Take "\textit{data\_directory}" as an example, \tool first splits it into \{"\textit{data}","\textit{directory}"\} and generates a regex expression as "\textit{data[\ .\_-]directory}". 
Then, \tool tries to match the regex expression with the log message.
If there is any match, \tool regards the log message as related to this configuration option.

However, If the name of a configuration option is too simple, 
such as configuration option "\textit{use}" in Nginx, there will be too many false positives by directly matching.
For a configuration option with a single word name \CodeIn{W}, 
\tool identifies whether \CodeIn{W} is a option name when matched in a log message by the following criteria:

\begin{itemize}
	\item[-] Is \CodeIn{W} quoted in the log message? (For example, the configuration option "\textit{alias}" in log message "\textit{`alias' cannot be used in location where URI was rewritten}")
	\item[-] Is there any specific modifier for \CodeIn{W} in the message that could indicate the identity. (namely "configuration", "config", "option", "directive", "parameter")
\end{itemize}

\subsubsection{\textbf{Finding Configuration Options implicitly related to Log Messages}}
\label{sec:design:select:implicit}

As mentioned in Section \ref{sec:study:analysis:relationship}, 
many log messages do not directly contain the information of the corresponding configuration option, 
but are influenced by variables and functions related to the option.
Consequently, if a log message doesn't match any option name explicitly,
\tool tries to find the options that are implicitly related to this log message
based on the summarized code patterns in Section \ref{sec:study:analysis:relationship}.

\vspace{3pt}
\romannumeral1) \textit{Collecting Related Code Elements of Configuration Options}

To identify whether a message is related to configuration options, the first thing is to figure out the code elements related to configuration options in source code.
Considering the summarized patterns of configuration-related log messages in Fig.\ref{fig:different-kinds-of-match-types}, \tool mainly focuses on the functions and variables related to configuration options. For ease of description, we refer to the variables related to configuration options as configuration variables, and functions related to configuration options as configuration functions.

Prior researchers\cite{xu2013not} observed that developers often use clean interface to manage the configuration-to-variable mapping information, namely structure, container, and comparison interface. 
\tool inherits and improves the method in \cite{zhou2016confmapper} to automatically collect program variables and functions related to configuration options.

\begin{figure*}
	\centering
	\includegraphics[width=.8\textwidth]{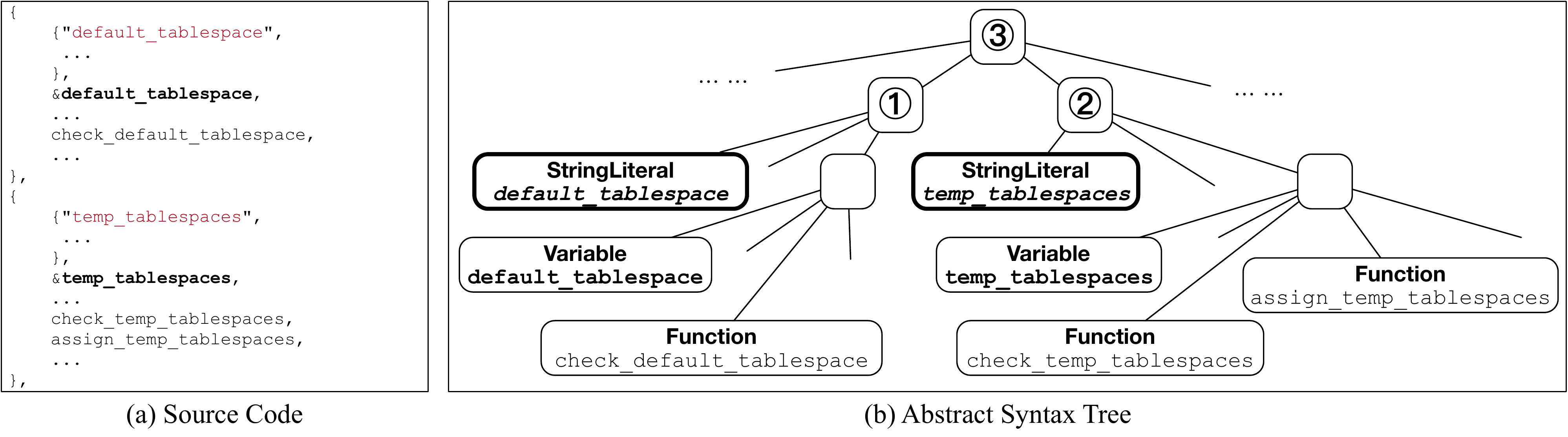}
	\vspace{-10pt}
	\caption{Example of collecting variables and functions based on minimal common ancestor algorithm.}
	\label{fig:ast-minimal-common-ancestor-example}
	\vspace{-10pt}
\end{figure*}

In detail, if the name string of configuration options are used in the structure interface defined for the configuration-variable mapping, 
\tool uses a minimal-common-ancestor algorithm to collect the configuration variables and configuration functions. 
The main idea is to find the minimal common subtree of the configuration option and its related code elements in AST (Abstract Syntax Tree), 
without interfering with other configuration options.
Take Fig.\ref{fig:ast-minimal-common-ancestor-example} as example, 
\tool finds that Node-\textcircled{\scriptsize 1} is the common ancestor of configuration option "\textit{default\_tablespace}" and its related code elements, then collects the related variables and functions as follows:

\indent \CodeIn{Variable}: "\textit{default\_tablespaces}"

\indent \CodeIn{Function}: "\textit{check\_default\_tablespaces}".

If the configuration options are mapped to variables with container interface (reading configuration value based on a getter api), 
\tool collects the \CodeIn{return variable} as configuration variable. 
Moreover, for comparison interface, 
if an configuration option is used in \CodeIn{strcmp}-like conditions, 
and the succeeding block is less than 3 lines, 
\tool collects the \CodeIn{left variable} of assign statement in the succeeding block as configuration variable.
That makes sense in most cases, 
since longer handling lines in succeeding block are more likely to contain other irrelevant objects.

There are situations that \tool could not collect any configuration variable during this procedure, 
for example, Nginx only declares the configuration functions in the structure interface, 
and assigns configuration variable with offset of structure pointers. \tool handles these situations in the \romannumeral3) step.

\vspace{3pt}
\romannumeral2) \textit{Backward Slicing on Log Messages}

Once the configuration variables and functions are collected, 
\tool applies a backward slicing on log messages to identify whether 
there is any configuration option related to this log message.

\tool begins the slicing from the statement that contains a log message. 
Once a variable or function is got in the slice, 
\tool checks whether it is a configuration variable or function, 
and regards the log message as related to this configuration option if it is. 
Take the code snippet in Fig.\ref{fig:motivating-example} as example, 
\tool first locates the log message in the \CodeIn{Return} statement, 
then orderly checks whether variable \CodeIn{map}, function \CodeIn{ap\_os\_is\_path\_absolute()} and \CodeIn{cmd->pool} are configuration variable or function.
Finally, when checking the function \CodeIn{vhost\_alias\_set()} in the slice, 
\tool finds that it is the configuration function of option "\textit{VirtualDocumentRoot}".
So \tool regards the log message "\textit{format string must be an absolute path, or `none'}" as related to "\textit{VirtualDocumentRoot}".

\vspace{3pt}
However, If the log message is in the structure outside the function declaration as shown below:

\begin{lstlisting}[language=myCinline]
AP_INIT_TAKE1("AddDefaultCharset", 
	set_add_default_charset, NULL, OR_FILEINFO,
	"The name of the default charset to add to any 
	Content-Type without one or 'Off' to disable"),
AP_INIT_TAKE1("AcceptPathInfo", 
	set_accept_path_info, NULL, OR_FILEINFO,
	"Set to on or off for PATH_INFO to be accepted by 
	handlers, or default for the per-handler preference"),
\end{lstlisting}

\tool traverses the parent subtree of this log message in AST until it meets the name string of any configuration option or reaches the top of current structure's subtree. 
In this example, \tool first locates the \CodeIn{StringLiteral} AST node 
containing log message "\textit{The name of the default charset to ...}" in line 3$\sim$4,
then traverses the parent subtree of current node, 
until reaches the subtree of structure \CodeIn{AP\_INIT\_TAKE1}, i.e. line 1$\sim$4.
During this procedure, \tool finds the option "\textit{AddDefaultCharset}", 
so \tool regards the log messages as related to "\textit{AddDefaultCharset}".

\vspace{3pt}
\romannumeral3) \textit{Complementary Similarity Comparison}

As mentioned in step \romannumeral1), 
there are situations that no configuration variable is collected by \tool. 
If there is either no matched configuration function in step \romannumeral2), 
\tool applies an approximate approach to identify configuration-related log messages based on the similarity between configuration options and the variables in the slice.

Developers usually tend to name the configuration variable similar to the option\cite{lawrie2006what,liu2016nomen,rice2017detecting,pandita2012inferring,raychev2014code,raychev2015predicting}.
Thus, \tool applies the Jaccard Index\cite{tan2016introduction} to calculate the similarity between variable and configuration option and infer whether they are relevant.
If a variable in the slice of a log message has a high similarity with one configuration option,
\tool regards this log message as related to the option.
In detail, after doing segmentation and lemmatization for configuration option and variable, 
\tool applies the \textit{idf}\cite{jones1972statistical} (inverse document frequency) to calculate the weight of every word, by taking each configuration option as a document, and all the configuration options as the corpus. We selected the \textit{idf} measure since it has been theoretically proved its effectiveness in weighting\cite{jones1972statistical} and practically applied in prior work\cite{zhang2015proactive}.
Then \tool calculates the similarity as follows:

\begin{equation*}
	\begin{split}
	Sim&(Conf\_Name,Var\_Name) =\\ 
	&\frac{\sum idf(word_i)~foreach~word_i \in Cwords \cap Vwords}{\sum idf(word_j)~foreach~word_j \in Cwords \cup Vwords}
	\end{split}
\end{equation*}

\begin{table*}[htb]
	\caption{POS Patterns of Constraints' Description.}
	\label{table:constraint-patterns}
	\centering
	\setlength{\tabcolsep}{1mm}{
		\begin{tabular}{lcl}
		\toprule
			\multicolumn{1}{c}{\textbf{Pattern}} &
			\textbf{Percentage.} &
			\multicolumn{1}{c}{\textbf{Example}} \\ 
		\midrule
			NN MD(must/should/...)$^\dagger$ [not]$^\ddagger$ VB(be/have/...)&	56.3\%	&	This value must be greater than 0.	\\
			NN VB(accept/require/...) NN				&	19.3\%	&	operationProfiling.slowOpSampleRate accepts values between 0 and 1.	\\
			JJ(acceptable/valid/...) NN VB(be)			&	14.7\%	&	Valid options are `ALWAYS', `NEVER', and `ONCE'.	\\
			NN VB(be) [not] JJ(acceptable/valid/...)	&	7.0\%	&	Negative value or 0 are invalid values and will fail NN startup.	\\
			NN(ERROR\_STATUS), NN(CONFIG) VB(be/have/...) JJ(less/greater/...)	& 	2.8\%	&	Error: olcIdleTimeout is less than 0.	\\
		\bottomrule
		\end{tabular}
		\begin{tablenotes}
			\item {$^\dagger$ () means one of them. $^\ddagger$ [] means optional}
		\end{tablenotes}
	}
	\vspace{-10pt}
\end{table*}

The \CodeIn{Cwords} and \CodeIn{Vwords} in the equation refer to the word set of configuration option and variable after segmentation and lemmatization correspondingly. 
To evaluate whether a variable is related to the configuration option, 
a threshold $\mu$ for the similarity is needed. 
We will evaluate the proper value in Section \ref{sec:evaluation:RQ2}.

\subsection{Constraints Inferring}
\label{sec:design:infer}

After the selection step, \tool obtains the configuration-related log messages.
Then, \tool mainly infers the constraints description from log messages.
In general, the log messages written in natural language are often unstructured and with various purposes, 
such as monitoring task progress, pinpointing error status, and guiding users' fixing. 
To infer the constraints in log messages, we applied a NLP (Natural Language Processing) pattern-based method to represent the description of configuration constraints. 
We studied and summarized the patterns of constraint description from 11 open-source software systems\footnotemark[1]. They are different from those we studied in Section \ref{sec:study} with different functionalities.

\footnotetext[1]{The 11 open-source software systems are Hadoop, HDFS, Yarn, Alluxio, Cassandra, Spark, Hypertable, MongoDB, AOLServer, Subversion, OpenLDAP.}

Two authors conducted the job to collect constraints description from documents and log messages.
For documents, they manually read the documents and located the description about the usage of each configuration option, 
then collected the text part that specify the constraints.
For log messages in source code, they located the log messages containing configuration options' name at first, 
then manually comprehended the context and meaning of the log message to judge whether it is a description of configuration constraint.
After the collection, the two authors cross checked the constraints' description. 
If there was a disagreement, a third author was consulted for additional discussion until consensus was reached. 
Then, we combined their results and replaced the configuration names with keyword "CONFIG" to avoid the interference of option names.
Finally, we collected 338 constraints' description in total.

In order to summarize the patterns from those descriptions, 
we used POS (Part-Of-Speech) tag sequences to represent the descriptions and mine the common sequences as our pattern.
Firstly we collected the words related to error from WordNet\cite{wordnet} to build a list $\lambda$. 
For each description of configuration constraints, 
we generated the POS tag sequences with spaCy\cite{spacy-homepage}. 
For example, the constraints' description in the motivating example of Fig.\ref{fig:motivating-example} is transformed as 
"\CodeIn{NN(format)  NN(string)  MD(must)  VB(be)  DT(an)  JJ(absolute)  NN(path)  ,  CC(or)  ``  NN(none)  ''  .}".
We removed the \CodeIn{DT}, \CodeIn{SYM} tags in tag sequences, 
and merged the successive same tags for \CodeIn{NN}, \CodeIn{JJ} as one.
If there was any word of a \CodeIn{NN} matched in $\lambda$, 
we kept the tag \CodeIn{NN} and replaced the word with "ERROR\_STATUS".
For instance, the above tag sequence is transformed as "\CodeIn{NN(format string) MD(must) VB(be) JJ(absolute) NN(path) CC(or) NN(none)}". 
Finally, we mined the common sequence among all the tag sequences, 
and found that 96.7\% of the descriptions fall into the patterns illustrated in Table \ref{table:constraint-patterns}.

When applying the patterns in recognizing log messages that describe configuration constraints,
\tool first replaces the configuration name with "CONFIG" if it exists in a log message, 
and replaces the word matched in $\lambda$ with "ERROR\_STATUS". 
Then \tool generates the corresponding POS tag sequence of the log message. 
Finally, if there is any match of the patterns illustrated in Table \ref{table:constraint-patterns}, \tool regards it as a configuration constraints.

\section{Evaluation}
\label{sec:evaluation}

To evaluate \tool, we try to answer three Research Questions as follows.

\textbf{RQ1}: How effective is \tool on inferring configuration constraints?

\textbf{RQ2}: How accurate is \tool on selecting configuration-related log messages?

\textbf{RQ3}: How does \tool compared with the state-of-the-art tool for inferring configuration constraints?

For RQ1, we evaluate the effectiveness of \tool in inferring configuration constraints.
For RQ2, we discuss how to set the parameter $\mu$ in \tool, 
and the final precision of \tool to select configuration-related log messages. 
For RQ3, we compared \tool with the state-of-the-art tool SPEX\cite{xu2013not},
and the dynamic method of configuration error injection\cite{zhang2015proactive}.

\subsection{Experiment Setup}

We conducted experiments on seven software systems listed in Table \ref{table:effectiveness} to evaluate \tool from different aspects. 
In case of the bias from our empirical study, 
we added three software systems to prove the generalization. 
All these seven software systems are widely-deployed, 
and representative across a number of system software.
We set the threshold $\mu$ during the selection of configuration-relation log messages as 0.63, 
and introduce how to set this threshold in Section \ref{sec:evaluation:RQ2}.

\begin{table}[tb]
	\caption{The Effectiveness of Inferring Configuration Constraints}
	\label{table:effectiveness}
	\centering
	\setlength{\tabcolsep}{3mm}{
		\renewcommand{\arraystretch}{1.2}
		\begin{tabular}{cccc}
		\toprule
			\multirow{2}{*}[-8pt]{\textbf{Software}} & \multicolumn{3}{c}{\textbf{\# of Config Constraints}} \\ \cline{2-4}
				&	\textbf{True Positive} & 
					\textbf{False Positive} &
					\textbf{False Negative}   \\
		\midrule										
			Httpd          & 164  	&	41	 & 	23 	\\	
			Nginx          & 80	    &	23	 &	8	\\	
			MySQL          & 33		&	9	 & 	10	\\	
			PostgreSQL     & 35    	&	6	 & 	7	\\	
			Lighttpd       & 63   	&	19	 & 	12	\\	
			Squid          & 30    	&	13	 & 	8	\\	
			Postfix        & 22	    &	8	 & 	12	\\ 	
		\midrule
			\textbf{Total}	& 	\textbf{427}	& \textbf{119}	& \textbf{80}	\\
		\bottomrule
		\end{tabular}
	}
	\vspace{-10pt}
\end{table}

\subsection{RQ1: Effectiveness of Inferring Configuration Constraints}
\label{sec:evaluation:RQ1}

To answer RQ1, we manually checked the results of \tool in inferring configuration constraints. 
The statistics is shown in Table \ref{table:effectiveness}. 
Generally, \tool could infer 22$\sim$164 configuration constraints from log messages, 
in which the average precision is 72.1\% (427/546), and the average recall is 18.7\%.

The false positives are mainly introduced by the incorrect selection of configuration-related log messages, 
and the mismatches of POS patterns in log messages not describing constraints. 
For instance, the log message "\textit{Symbolic link is not allowed or link target not accessible}" is mis-matched with configuration option "\textit{AcceptPathInfo}" due to the wrong configuration variable \CodeIn{r->path\_info}, and this log message is also mistakenly inferred as an constraint of "\textit{AcceptPathInfo}".
The main reason for the false negatives is that the descriptions of these constraints are mainly describing the error status of configuration, 
which could not match our POS patterns, but can be inferred by manual comprehension. 
For instance, the log message "\textit{auth.backend.ldap.filter is missing a replace-operator `?'}" of option "\textit{auth.backend.ldap.filter}" in Lighttpd indicates that there should be a "\textit{?}", which could not be matched with any of the POS patterns.

\begin{table}[tb]
	\caption{Submitted Patches to enhance documentation.}
	\label{table:patch-info}
	\centering
	\setlength{\tabcolsep}{3mm}{
		\begin{tabular}{cccc}
		\toprule
		\textbf{Software} & 
		\textbf{\begin{tabular}[c]{@{}c@{}}\# of Submitted\\Patches$^\dagger$\end{tabular}} &
		\textbf{\begin{tabular}[c]{@{}c@{}}\# of Confirmed\\Patches\end{tabular}} &
		\textbf{\begin{tabular}[c]{@{}c@{}}\# of Accepted\\Patches\end{tabular}}  \\ 
		\midrule
			Httpd      & 	25 &	4 	&	3	\\	\Space{fixed: 64893,64904,64909; wontfix: 64897}
			Nginx      & 	14 & 	13 	&	0	\\	\Space{wontfix: }
			MySQL      & 	7  & 	7	&	3	\\	\Space{verified: 101512,101515,101516,101519; fixed: 101513,101514,101520}
			PostgreSQL & 	2  & 	0	&	0	\\	
			Lighttpd   & 	4  & 	4	&	4	\\	\Space{fixed: 3035,3036,3038,3040}
			Squid      & 	8  & 	1	&	0	\\	\Space{5095}
			Postfix    & 	7  & 	0	&	0	\\
		\midrule
			\textbf{Total}	& \textbf{67}	& \textbf{29} &  \textbf{10}\footnotemark[2]  \\
		\bottomrule
		\end{tabular}
	}
	\vspace{-10pt}
\end{table}

\footnotetext[2]{Accepted bug IDs: Httpd-\{64893, 64904, 64909\}, MySQL-\{101513, 101514, 101520\}, Lighttpd-\{3035, 3036, 3038, 3040\}.}

To prove the effectiveness of inferring configuration constraints from source code by \tool, 
we manually checked whether the constraints are recorded in the documentation. 
As shown in Table \ref{table:patch-info}, 
we finally submitted 67 patches to enhance the documentation about description of configuration constraints.
The constraints in 29 patches have been confirmed by the developers, among which 10 patches have been accepted. The other 38 patches are still pending for responses at the time of writing.
For the 19 confirmed patches that have not been accepted, 4 patches are waiting for developers' update in documentation, 
and the other 15 patches are not accepted mainly due to developers' concern about complicating the documentation. 
Nevertheless, these constraints could be applied to guide the checking of misconfiguration in production.

\subsection{RQ2: Precision of Selecting Configuration-Related Log Messages}
\label{sec:evaluation:RQ2}

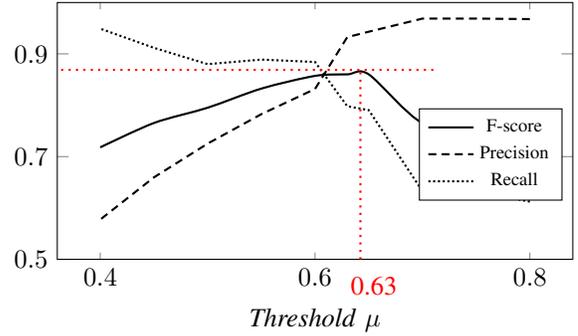
\begin{figure}
	\centering
	\begin{tikzpicture}{}
		\begin{axis}[
				ymin=0.5, ymax=1,	
				xlabel=\textit{Threshold $\mu$},
				xtick={0.4, 0.6, 0.8},
				ytick={0.5, 0.7, 0.9},	
				xscale=1, 
				yscale=0.6,
			]
			\legend{\scriptsize{F-score}, \scriptsize{Precision}, \scriptsize{Recall}}
			\addplot[smooth, thick]	
			coordinates{
				( 0.8, 0.750000) (0.75, 0.765432) ( 0.7, 0.765432) (0.65, 0.860104) (0.63, 0.860104) 
				( 0.6, 0.857143) (0.55, 0.832370) ( 0.5, 0.795181) (0.45, 0.765432) ( 0.4, 0.717949)
			};	
			\addplot[densely dashed, thick]	
			coordinates{
				( 0.8, 0.967742) (0.75, 0.968750) ( 0.7, 0.968750) (0.65, 0.943182) (0.63, 0.932584) 
				( 0.6, 0.831683) (0.55, 0.782609) ( 0.5, 0.725275) (0.45, 0.659574) ( 0.4, 0.577320)
			};	
			\addplot[densely dotted, thick]	
			coordinates{
				( 0.8, 0.612245) (0.75, 0.632653) ( 0.7, 0.632653) (0.65, 0.790476) (0.63, 0.798077) 
				( 0.6, 0.884211) (0.55, 0.888889) ( 0.5, 0.880000) (0.45, 0.911765) ( 0.4, 0.949153)
			};	
		\end{axis}
		\draw [red, dotted, thick] (4.03, 0)--(4.03, 2.52);
		\draw [red, dotted, thick] (5, 2.52)--(0, 2.52);
		\draw (4.2, 0) -- (4.21, 0) node[below=3pt, color=red]{0.63};
	\end{tikzpicture}
	\caption{F1-score of different threshold $\mu$ in calculating similarity between configuration options and program variables.}
	\label{fig:similarity-threshold}
\end{figure}

\begin{table}[tb]
	\caption{Precision of \tool in Selecting Configuration-related Log Messages.}
	\label{table:select-precision}
	\centering
	\setlength{\tabcolsep}{5mm}{
		\begin{tabular}{ccc}
		\toprule
			\textbf{Software} &
			\textbf{\begin{tabular}[c]{@{}c@{}}\# of Config-related \\Log Messages\end{tabular}} &
			\textbf{\begin{tabular}[c]{@{}c@{}}Precision\\(in 100 samples)\end{tabular}} \\ 
		\midrule
			Httpd          & 1545   &	72.4\%	\\
			Nginx          & 773    &	78.1\%	\\
			MySQL          & 1097   &	73.5\%	\\
			PostgreSQL     & 752    &	74.7\%	\\
			Lighttpd       & 396    &	65.9\%	\\
			Squid          & 333    &	63.1\%	\\
			Postfix        & 470    &	68.5\%	\\ 
		\midrule
			\textbf{Total}	&	\textbf{5366} & \textbf{70.7\%} \\
		\bottomrule
		\end{tabular}
	}
	\vspace{-10pt}
\end{table}

\begin{table*}[tb]
	\caption{Comparison with the State-of-the-art.}
	\label{table:comparison}
	\centering
	\setlength{\tabcolsep}{1mm}{
		\renewcommand{\arraystretch}{1.2}
		\begin{tabular}{cccccc}
			\toprule
			\textbf{Software} &
			  \textbf{\begin{tabular}[c]{@{}c@{}}Constraints that \\SPEX Could Infer\end{tabular}} &
			  \textbf{\begin{tabular}[c]{@{}c@{}}Constraints Inferred by \tool \\While SPEX Could \textit{NOT} Infer\end{tabular}} &
			  \textbf{\begin{tabular}[c]{@{}c@{}}Constraints Inferred by \tool \\While Dynamic Method Could \textit{NOT} Infer\end{tabular}}  & 
			  \textbf{\begin{tabular}[c]{@{}c@{}}Constraints that Could Only\\be Inferred by \tool\end{tabular}} \\ 
			\midrule
				\textbf{Httpd}		& 78	& 88 (53.7\%)	&	46 (28.0\%)	 &	41 (25.0\%) \\
				\textbf{Nginx}		& 182	& 42 (51.3\%)	&	73 (91.3\%)	 &	39 (48.8\%) \\
				\textbf{MySQL}		& 351	& 17 (51.5\%)	&	30 (90.9\%)	 &	15 (45.5\%) \\
				\textbf{PostgreSQL}	& 263	& 27 (85.7\%)	&	26 (74.3\%)	 &	19 (54.3\%) \\
				\textbf{Lighttpd}	& 95	& 48 (76.2\%)	&	48 (76.2\%)	 &	37 (58.7\%) \\
				\textbf{Squid}		& 189	& 22 (73.3\%)	&	-$^\dagger$	 &	- \\
				\textbf{Postfix}	& 137	& 19 (86.4\%)	&	-$^\dagger$	 &	- \\ 
			\midrule
				\textbf{Total}		&	\textbf{1295}	& \textbf{263 (61.6\%)}	&	\textbf{223 (59.5\%)}	\Space{223/375}	& \textbf{151 (40.3\%)} \Space{151/375} \\
			\bottomrule
		\end{tabular}
		\begin{tablenotes}
			\item {$^\dagger$ Don't have official test suite.}
		\end{tablenotes}
	}
	\vspace{-10pt}
\end{table*}

To answer RQ2, we evaluated the precision of \tool in selecting Configuration-Related Log Messages. 
During the procedure of \tool, 
the threshold $\mu$ of similarity between configuration option and variable related to log message requires to be predefined. 
We first tune the threshold $\mu$, then evaluate the precision.

To determine the proper value of threshold $\mu$, 
we randomly sampled 100 configuration options from Httpd and Nginx, each for 50, 
and manually inspected the corresponding program variables in source code as oracle. 
If the option doesn't have a corresponding variable, 
like the structural option "<Directory>" in Httpd, 
we sample another option. 
Then, we manually set the threshold $\mu$ to different value, 
and run \tool to find the corresponding variables for the 100 configuration options. 
Finally, we verified the variables found by \tool with the oracle, 
and calculated the F1-score by $\frac{2*Precision*Recall}{Precision+Recall}$. 
Fig.\ref{fig:similarity-threshold} shows the F1-score with different value of threshold $\mu$, 
and when $\mu$ = 0.63, F1-score reaches the highest value. 
Based on the result, we set threshold $\mu$ = 0.63 in \tool.

We ran \tool on seven software systems to select the log messages that related to configuration options. To evaluate the precision, we randomly sampled 100 log messages for each software system to inspect whether the log message is really related to the specific configuration options. The statistics shown in Table \ref{table:select-precision} illustrate the precision of configuration-related log messages selection.

\subsection{RQ3: Comparison with the State-of-the-art}
\label{sec:evaluate:RQ3}

\subsubsection{Comparison with SPEX}
\label{sec:evaluate:RQ3:spex}

We compared \tool with SPEX, which is one of the most effective constraint inferring tools. 
Follow the design of SPEX, we tried our best to annotate the mapping interfaces in source code, 
and implemented the constraints extraction.
The results are shown in the third column of Table \ref{table:comparison}.
SPEX mainly infers configuration constraints based on searching code patterns for different predefined constraints types. 
Though the total amount of configuration constraints that SPEX could infer is relative large as shown in the second column,
\tool could still infer up to 61.6\% (263/427) of the constraints that SPEX could not.

Among the constraints that could not be inferred by SPEX, 
55.1\% (145/263) are descriptions about constraints related to the semantic information of complicated context. 
For example, the constraint "\textit{binlog\_checksum should be NONE for Group Replication}" in MySQL, 
and "\textit{the `alias' directive cannot be used inside the named location}" in Nginx, all include the summary of context information. 
35.7\% (94/263) are descriptions about constraints related to the syntax format of configuration values. 
For example, the configuration option "\textit{server.modules}" in Lighttpd requires "\textit{list of `mod\_xxxxxx' strings}". 
The remaining 9.1\% (24/263) are various causes. 
For example, the value range of configuration option "\textit{MaxRangeOverlaps}" in Httpd is a combination of enumerated and numeric values, 
requiring "\textit{`none', `default', `unlimited' or a positive integer}", 
which is not covered by SPEX's predefined constraint types; 
for configuration option "\textit{old\_passwords}" in MySQL, 
the valid values are "2" and "0" without "1", 
but SPEX can only infer that it could not be "1" and didn't know what is the valid values.

\subsubsection{Comparison with Dynamic Method}
\label{sec:evaluate:RQ3:dynamic}

As mentioned in Section \ref{sec:study:result}, 
we applied a mutation based configuration error injection method, 
which we refer to as dynamic method here, 
on studied software systems, 
and obtained a few configuration constraints from log records. 
We compared \tool with dynamic method as follows. 
Considering that Squid and Postfix don't provide official test suite, 
we only conduct the dynamic method on other five software systems.
We applied the mutation based method to generate configuration errors for the configuration options that \tool could infer constraints, 
then ran the test suite to obtain log records and check whether there is any constraints description about the configuration options.
The detailed statistics are listed in the fourth column of Table \ref{table:comparison}. 
On average, 59.5\% (223/375) of configuration constraints inferred by \tool could not be triggered by dynamic method.

We further analyzed the causes, and found the reasons are mainly threefold.
First, the configuration values often have syntax requirements. 
For example, 31 options in Lighttpd require a list of values, expressed as "(..., ...)". 
When generating a random string for these configuration options, 
the program output a log record as "\textit{should have been a array of strings like ... = ( ``...'')}", 
and only when the configuration value is in the required format with a wrong value, 
the program could indicate the constraints like "\textit{expected list of "path" strings}" in the log record. 
Second, the constraints that describe the dependency between multiple configuration options are difficult to trigger when generating configuration error for single option. 
For instance, the configuration option "\textit{proxy\_busy\_buffers\_size}" in Nginx must be less than the size of all "\textit{proxy\_buffers}" minus one buffer, 
this constraint was obtained by dynamic method in our experiment, 
while a similar option pair "\textit{fastcgi\_busy\_buffers\_size}" and "\textit{fastcgi\_buffers}" didn't trigger the relevant log statements, 
because randomly changing value of "\textit{fastcgi\_busy\_buffers\_size}" cannot ensure to break the relationship.
Last but not least, the various workload and limited test cases prevent the triggering of log records. 
For instance, in PostgreSQL, only when trying to make an online backup, 
the program checks the configuration option "\textit{wal\_level}" and indicates that the WAL level is not sufficient, 
which must be set to \CodeIn{replica} or \CodeIn{logical} at server start. 
As for the test cases, 
we ran the official test suite of the former five software system in Table \ref{table:comparison} and measured the coverage with \CodeIn{gcov}.
The statistics show that the code coverage could only reaches 19.2\%$\sim$65.1\%.

\subsubsection{Comparison with the Combination}
\label{sec:evaluate:RQ3:combination}

Finally, we compared the capability of \tool with the combined results of SPEX and dynamic method. 
As shown in the last column of Table \ref{table:comparison}, 40.3\% of the constraints that inferred by \tool could neither be inferred by SPEX nor dynamic method. This emphasizes that \tool could be an effective complementary to existing methods.

\section{Threats to Validity}
\label{sec:threats}

\textbf{Quality of log messages:} \tool infers configuration constraints from log messages in source code. The process works under the premise that the log messages contain descriptions of constraints. We believe that logs written by experienced developers in large-scale software systems usually contain description about system status or guidance for the failure diagnosis. In such a setting, ConfInLog could infer configuration constraints from log messages.

\textbf{Generalization of summarzied patterns:} \tool selects the configuration-related log messages and infers constraints from log messages based on summarized patterns. The generalization of those patterns influence the effectiveness of \tool. To cover different situations as many as possible, we chose different kinds of software systems to summarize the patterns of configuration-related log messages. As for the nlp patterns of constriants' description, to avoid over-fitting of our studied software systems, we chose 11 software systems different from the evaluated ones to summarize the patterns.

\section{Related Work}
\label{sec:rw}

\subsection{Misconfiguration Detection}
The prevalence and severity of software misconfigurations have promoted the development of many diagnosis 
and detection technologies \cite{Attariyan2010Automating, Attariyan2012XRay, Rabkin2011Precomputing, zhang2013automated, bauer2011detecting,xu2013not, Zhang2014EnCore, Li2017ConfTest, xiang2020pracextractor, Yuan2011, Attariyan2008, Zhen2015, Huang2015ConfValley, santolucito2017synthesizing, tang2015holistic, xu2016early, zhang2015proactive, Xiang2019}.
Misconfiguration diagnosis refers to locate root causes of misconfigurations that caused failures, performance issues, 
and incorrect behaviors, while detection aims at detecting misconfigurations before deployment.
Most of the detection work discovers configuration errors by either checking against (a) rules mined from numbers of 
configuration files and system behaviors, or (b) constraints inferred from source code and documentation. Our 
work targets at inferring configuration constraints from the log messages in source code, in order to prevent and detect 
misconfigurations.

EnCore \cite{Zhang2014EnCore} uses machine-learning method to infer the configuration rules between configuration settings 
and the executing environment for misconfiguration detection.
ConfigV \cite{santolucito2017synthesizing} learns specification from configuration files based on the association rule learning algorithm to detect ordering errors, integer correlation errors, type errors, and missing entry errors.
CODE \cite{Yuan2011} exploits the registry's access rules in Windows to detect configuration errors by finding violations of the rules.
These approaches rely on the correctness and completeness of the learned samples.

Prior work\cite{xu2013not,rabkin2011static,chen2020understanding,liao2018you} proposes static analysis approaches to infer configuration constraints from source code.
SPEX\cite{xu2013not} searches code patterns for different predefined constraints types, such as data range and value relationship. 
CDep\cite{chen2020understanding} focuses on five types of configuration dependency between multiple 
configuration parameters.
However, these researches are limited by the predefined code patterns 
and accuracy of program analysis.
ConfTest\cite{Li2017ConfTest} summarizes a fine-grained classification of option types and infers syntactic and 
semantic constraints of each type based on the classification.
PracExtractor\cite{xiang2020pracextractor} employs Natural Language Processing techniques to automatically infer configuration specifications from software manuals. 
Our work mainly focus constraints in configuration-related log messages in source code,
which is complementary to the prior work.

\subsection{Analyzing Logs}
Logs contain rich information which
helps people understand the running state of
the program.
Prior work on analyzing logs mainly focus on 
debugging failures and performance problems.

The literature~\cite{Yuan2010SherLog, beschastnikh2014inferring,
xu2009detecting, mariani2008automated, lou2010mining,
zhao2017log20, zhao2014lprof, huang2018capturing} uses different methods
to diagnose functional system failures and performance faults 
using system log information. For instance,
SherLog~\cite{Yuan2010SherLog} analyzes 
the logs for deterministic replay in conjunction 
with source code to locate the execution
paths and contexts when software errors occur. 
Lprof~\cite{zhao2014lprof} associates dispersed log entries to specific 
individual requests and automatically reconstructs 
the execution flow of each request to detect 
performance anomalies. Similarly,
Panorama~\cite{huang2018capturing} leverages
logs to observe interactions between a system's components
to find out the gray failures.

Our study mainly focuses on analyzing logs to
infer configuration constraints
inside the log messages. To the best of our knowledge,
few research has been conducted before.

\section{Conclusion}
\label{sec:conclusion}

Given that software logs often contain configuration constraints,
this paper conducted an empirical study and summarized several patterns of configuration-related log messages.
Guided by the study, we designed \tool, a static tool to infer configuration constraints from log messages. 
An experiment on seven popular open-source software systems proves its effectiveness. 
\tool could infer 22$\sim$163 constraints for the software systems, 
in which 59.5\%$\sim$61.6\% could not be inferred by the state-of-the-art work.
Finally, we submitted 67 documentation patches regarding the constraints inferred by \tool. The constraints in 29 patches have been confirmed by the developers, among which 10 patches have been accepted.
\section*{Acknowledgement}
\label{sec:ack}

This research was substantially supported by National Key R\&D Program of China (Project No. 2017YFB1001802); 
National Natural Science Foundation of China (Project No. 61872373 and No. 61872375).


\bibliographystyle{IEEEtran}
\bibliography{refs}

\end{document}